\documentclass[twocolumn,prl,floatfix]{revtex4} \pdfoutput=1
\usepackage{graphicx} \usepackage{amssymb} \usepackage{amsmath}
\usepackage{tensor} 

\begin{document}

\makeatletter \def\imod#1{\allowbreak\mkern10mu({\operator@font
    mod}\,\,#1)} \makeatother

\newcommand{\eg}{{\em e.g.}}
\newcommand{\ie}{{\em i.e.}}
\newcommand{\cf}{{\em cf.}}
\newcommand{\N}{\mbox{\boldmath{$N$}}}
\newcommand{\eqn}[1]{(\ref{eq.#1})}
\newcommand{\dotprod}{{\scriptscriptstyle \stackrel{\bullet}{{}}}}
\newcommand{\bra}[1]{\mbox{$\left\langle #1 \right|$}}
\newcommand{\ket}[1]{\mbox{$\left| #1 \right\rangle$}}
\newcommand{\bl}{{\mbox{\rm \scriptsize B}}}
\newcommand{\blket}[1]{\mbox{$\left| #1 \right\rangle_\bl$}}
\newcommand{\braket}[2]{\mbox{$\left\langle #1 | #2 \right\rangle$}}
\newcommand{\blbraket}[2]{\mbox{$\left\langle #1 | #2\right\rangle_\bl$}}
\newcommand{\av}[1]{\mbox{$\left| #1 \right|$}}
\newcommand{\bk}[1]{\mbox{$\left\langle #1 \right\rangle$}}
\newcommand{\ve}{\varepsilon}
\newcommand{\osc}{{\mbox{\rm \scriptsize osc}}}
\newcommand{\tot}{{\mbox{\rm \scriptsize tot}}}
\newcommand{\op}[1]{\mathsf{#1}} \newcommand{\tts}{{\hspace{.1em}}}
\newcommand{\lga}{{\mbox{\rm \scriptsize LGA}}}
\newcommand{\swap}{{\mbox{\rm \scriptsize swap}}}
\newcommand{\ground}{{\mbox{\rm \scriptsize ground}}}
\newcommand{\cycle}{{\mbox{\rm \scriptsize cycle}}}
\newcommand{\orbit}{{\mbox{\rm \scriptsize orbit}}}
\newcommand{\interval}{{\mbox{\rm \scriptsize interval}}}
\newcommand{\diff}{{\mbox{\rm \scriptsize diff}}}
\newcommand{\freelymoving}{{\mbox{\rm \scriptsize freely moving}}}
\newcommand{\massless}{{\mbox{\rm \scriptsize massless particles}}}
\newcommand{\light}{{\mbox{\rm \scriptsize light}}}
\newcommand{\particle}{{\mbox{\rm \scriptsize particle}}}
\newcommand{\particles}{{\mbox{\rm \scriptsize particles}}}
\newcommand{\classical}{{\mbox{\rm \scriptsize classical}}}
\newcommand{\internal}{{\mbox{\rm \scriptsize internal}}}
\newcommand{\initial}{{\mbox{\rm \scriptsize initial}}}
\newcommand{\midpoint}{{\mbox{\rm \scriptsize midpoint}}}
\newcommand{\average}{{\mbox{\rm \scriptsize average}}}
\newcommand{\nonrel}{{\mbox{\rm \scriptsize non-rel}}}
\newcommand{\rest}{{\mbox{\rm \scriptsize rest}}}
\newcommand{\twoparticles}{{\mbox{\rm \scriptsize 2p}}}
\newcommand{\ideal}{{\mbox{\rm \scriptsize ideal}}}
\newcommand{\block}{{\mbox{\rm \scriptsize block}}}
\newcommand{\blockchange}{{\mbox{\rm \scriptsize block-change}}}
\newcommand{\statechange}{{\mbox{\rm \scriptsize state-change}}}
\newcommand{\changed}{{\mbox{\rm \scriptsize changed}}}
\newcommand{\unchanged}{{\mbox{\rm \scriptsize unchanged}}}
\newcommand{\change}{{\mbox{\rm \scriptsize change}}}
\newcommand{\nochange}{{\mbox{\rm \scriptsize same}}}
\newcommand{\kinetic}{{\mbox{\rm \scriptsize kinetic}}}
\newcommand{\potential}{{\mbox{\rm \scriptsize potential}}}
\newcommand{\motion}{{\mbox{\rm \scriptsize motion}}}
\newcommand{\local}{{\mbox{\rm \scriptsize local}}}
\newcommand{\hop}{{\mbox{\rm \scriptsize hop}}}
\newcommand{\locations}{{\mbox{\rm \scriptsize locations}}}
\newcommand{\even}{{\mbox{\rm \scriptsize even}}}
\newcommand{\odd}{{\mbox{\rm \scriptsize odd}}}
\newcommand{\rt}{{\mbox{\rm \scriptsize R}}}
\newcommand{\lt}{{\mbox{\rm \scriptsize L}}}
\newcommand{\shift}{{\mbox{\rm \scriptsize shift}}}
\newcommand{\ex}{{\mbox{\rm \scriptsize ex}}}
\newcommand{\D}[2]{\frac{\partial #2}{\partial #1}}
\newcommand{\pp}{{\mbox{\tt \scriptsize /}}}
\newcommand{\mm}{{\mbox{\tt \scriptsize \backslash}}}
\newcommand{\x}{{\tilde{x}}} \newcommand{\y}{{\tilde{y}}}
\newcommand{\z}{{\tilde{z}}} \newcommand{\tn}{{\tilde{t}}}
\newcommand{\xb}{{\bar{x}}}
\newcommand{\smax}{{\mbox{\rm \scriptsize max}}}
\newcommand{\smin}{{\mbox{\rm \scriptsize min}}}
\newcommand{\ensemble}{{\mbox{\rm \scriptsize ensemble}}}

\newcommand{\taumin}{\tau_{\smin}} \newcommand{\tauavg}{\tau}

\newcommand{\n}{{\boldsymbol{n}}} \newcommand{\m}{{\boldsymbol{m}}}
\renewcommand{\i}{{\boldsymbol{i}}} 
\renewcommand{\j}{{\boldsymbol{j}}} 
\renewcommand{\k}{{\boldsymbol{k}}} 
\renewcommand\Re{\operatorname{\mathfrak{Re}}}

\newcommand{\sinc}{{\mbox{\rm sinc}}\:}
\newcommand{\sincs}{{\mbox{\rm sinc$^2$}}}

\newcommand{\figw}[3]{
  \begin{figure*}
    $$#2$$\relax
    \caption{#3}
    \label{fig.#1}
\end{figure*}                 }

\newcommand{\fign}[3]{
  \begin{figure}
    $$#2$$\relax
    \caption{#3}
    \label{fig.#1}
\end{figure}                 }

\setlength{\fboxsep}{.1pt} \setlength{\fboxrule}{.1pt}


\title{Quantum emulation of classical dynamics}

\author{Norman Margolus}
\email{nhm@mit.edu}
\affiliation{Massachusetts Institute of Technology}

\date{\today}




\begin{abstract} 


  {\em In statistical mechanics, it is well known that finite-state
    classical lattice models can be recast as quantum models, with
    distinct classical configurations identified with orthogonal basis
    states.  This mapping makes classical statistical mechanics on a
    lattice a special case of quantum statistical mechanics, and
    classical combinatorial entropy a special case of quantum entropy.

    In a similar manner, finite-state classical dynamics can be recast
    as finite-energy quantum dynamics.  This mapping translates
    continuous quantities, concepts and machinery of quantum mechanics
    into a simplified finite-state context in which they have a purely
    classical and combinatorial interpretation.  For example, in this
    mapping quantum average energy becomes the classical update rate.

    Interpolation theory and communication theory help explain the
    truce achieved here between perfect classical determinism and
    quantum uncertainty, and between discrete and continuous
    dynamics.}




\end{abstract}

\keywords{mechanics, bandlimited, interpolation, uncertainty-relations}

\maketitle


\section{Introduction}\label{sec.intro}

In this paper we discuss a mapping between classical and quantum
systems that lets us regard quantum dynamics as a generalization of
finite state classical dynamics, and that allows us to identify
equivalent quantities and concepts in classical and quantum systems.



A similar mapping has long been known in statistical mechanics
\cite{ruelle} that establishes classical lattice models and their
combinatorial entropy as simple examples of quantum statistical
mechanics.

There is an obvious candidate for the comparable dynamical
mapping: classical computations are equivalent to a subset of
quantum computations \cite{bennett-info}.  Most work on quantum
computation is, however, based on hybrid classical/quantum models in
which macroscopic classical operations control the sequencing of
quantum operations.  Such systems do not provide a purely quantum
target for a classical/quantum mapping.  Instead, early work showing
that autonomous quantum systems can perform classical computation
\cite{benioff} forms the basis for the dynamical mapping presented
here.

This mapping allows physical quantities such as energy and momentum to
be identified with finite-state classical quantities, with the aid of
classical interpolation theory.  Related issues are addressed in
\cite{bcq}, but a general dynamical mapping is not provided there.

As a preliminary to discussing dynamics we first review a canonical
method for mapping classical lattice models onto quantum lattice
models in statistical mechanics.


\section{Statistical mechanics}\label{sec.stat-mech}

In statistical mechanics, it is well known that classical lattice
models can be recast as quantum models, with distinct classical
configurations identified with orthogonal basis states \cite{ruelle}.

Consider, for example, the well known ferromagnetic 2D Ising model.
In this model each of $M$ lattice sites in a square lattice is
occupied by a classical two-state ``spin,'' and each state $S_n$ of
the $N=2^M$ possible configurations of the lattice is assigned a
classical configurational energy $E_n^\classical$ that depends only on
how many pairs of adjacent lattice sites have the same spin value and
how many have opposite values.

A quantum lattice model corresponding to such a classical lattice
model can be constructed by identifying each of the $N$ distinct
classical states $S_n$ with a distinct basis vector $\ket{n}$ in an
$N$ dimensional Hilbert space.  A hamiltonian operator $\op{H}$ is
defined by taking each configuration state $\ket{n}$ to be an energy
eigenstate of $\op{H}$ with energy eigenvalue $E_n^\classical\;$:
\begin{equation}\label{eq.eclassical}
  \op{H}\,\ket{n}=E_n^\classical\,\ket{n}\;.
\end{equation}

In quantum statistical mechanics the energy eigenstates are also
eigenstates of the density operator $\op{\rho}$, with eigenvalues that
give the statistical weight to attach to each energy eigenstate.  For
example, for a canonical ensemble of quantum mechanical systems,
$\op{\rho}$ is proportional to $e^{-\beta \op{H}}$.  From
\eqn{eclassical} this becomes the usual classical Boltzmann factor
when applied to a configuration state $\ket{n}$, and quantum
statistical mechanics reduces to classical.

\section{Classical dynamics}

Since the definition of $\op{H}$ used in the statistical mechanics
mapping makes each classical configuration a time-invariant state
under unitary time evolution, we use a different definition of
$\op{H}$ to emulate classical dynamics.

\subsection{Finite-state dynamics}

An invertible classical finite-state dynamics is a discrete
idealization of classical dynamics \cite{fredkin-bbm}.  Perfect
digital degrees of freedom are updated at discrete times according to
a sequence of invertible transformations.  The total amount of state
in the system, including that used to define the dynamics, is finite.
Here we take the time between update events to always be $\tau$, so
that the system is updated at the constant rate $\nu=1/\tau$.

The finite set of possible configurations of the system is partitioned
by the invertible dynamics into a collection of disjoint {\em
  dynamical orbits}, with each dynamical orbit consisting of a set of
configurations that turn into each other under the dynamics
(\cf~\cite{toffoli-topics,coppersmith}).  For each dynamical orbit $d$
the number of configurations $N_d$ in the orbit determines the period
$T_d=\tau N_d$ of the orbit.  One configuration of each orbit is
labeled with the integer $0$.  The configuration obtained from $0$ by
one update step is labeled $1$, and so on.

We identify configuration $n$ of dynamical orbit $d$ with the basis
state $\ket{n,d}$.  Because the orbit is periodic,
$\ket{0,d}=\ket{N_d,d}$.  This mapping identifies each possible
configuration of the classical dynamics with a basis state: we call
this the configuration basis.

\subsection{Hamiltonian dynamics}\label{sec.hamiltonian}

Given an invertible classical finite-state dynamics, we construct a
continuous quantum hamiltonian dynamics isomorphic to the classical
dynamics at regularly-spaced times.  We begin by defining a discrete
Fourier transformed set of basis states.  Let
\begin{equation}\label{eq.Em}
  \ket{E\! :m,d} = {1\over \sqrt{N_d}}\sum_{n=0}^{N_d -1} e^{2\pi i n
    m/N_d}\,\ket{n,d}
\end{equation}
for integer $m$, where ``$E\,$'' is the name of the new basis.  The
inverse transformation is
\begin{equation}\label{eq.inv-trans}
  \ket{n,d} = {1\over \sqrt{N_d}}\sum_{m=0}^{N_d -1} e^{-2\pi i n
    m/N_d}\,\ket{E\! :m,d}\;.
\end{equation}
We define a hamiltonian $\op{H}$ by assigning the $\ket{E\! :m,d}$
states to be its energy eigenstates and $E_{m,d}=m\,h/T_d$ to be the
corresponding energy eigenvalues \footnote{This is equivalent to phase
  and number \cite[Eqn.~41]{pb}, with $\theta\,\ket{n,d}=(2\pi
  n/N_d)\,\ket{n,d}$ and $\mathbb{N}\,\ket{E\!  :m,d}=m\,\ket{E\!
    :m,d}$.}:
\begin{equation}\label{eq.eigen}
  \op{H}\,\ket{E\! :m,d} = m {h\over T_d}\,\ket{E\! :m,d}\;.
\end{equation}
If we let $\op{U}=e^{-i\op{H}\tau/\hbar}$ be the time evolution operator
for the time interval $\tau$, then
\begin{align}\label{eq.step}
  \op{U}\,\ket{n,d} &= {1\over \sqrt{N_d}}\sum_{m=0}^{N_d -1} e^{-2\pi
    i (n+1) m/N_d}\,\ket{E\! :m,d} \nonumber \\[.5em] &=
  \ket{n+1,d}\;.
\end{align}

\subsection{Average energy}

The configuration state \eqn{inv-trans} is a uniform superposition of
all $N_d$ energy eigenstates $\ket{E\! :m,d}$ with eigenvalues
$mh/T_d$, and so the average energy is
\begin{equation}\label{eq.EN}
  E={h(N_d - 1)\over 2T_d}\;.
\end{equation}
We've taken $E_{0,d}=0$ in the construction above, but the fact that
the system has a harmonic-oscillator-like energy spectrum suggests
that we should really add $h/2T_d$ to all the energy eigenvalues.
This is in fact the smallest energy allowed by quantum
distinguishability bounds, assuming the ground state energy of a much
larger system encompassing this one sets the zero of the energy scale
\cite{first-mom}.  Adding $h/2T_d$ makes the average energy \eqn{EN}
independent of $T_d$,
\begin{equation}\label{eq.E}
  E={h \nu \over 2}\;.
\end{equation}
This is the least possible average energy compatible with a dynamics
that traverses distinct states at the average rate $\nu$
\cite{first-mom}.  Thus our construction is energetically ideal, and
the average energy is identified with the classical update rate of the
finite-state dynamics.

If a lattice dynamics is updated sequentially---one location at a time
in a repeating cycle---the frequency with which a given location is
updated determines a local energy.  Total update frequency (total
energy) is the sum of the local frequencies \footnote{In this example
  the local additivity of the energy derives from the locality of the
  sequence of update operations; $\op{H}$ doesn't need to be
  expressible in a manifestly local fashion
  (\cf~\cite{thooft,marg-pqc}).}.  Different kinds of updates (\eg,
ones involving particle or bond motion, and ones that don't
\cite{3dlga,creutz}) define different kinds of energy \cite{bcq}.

For a large system with a very long period, $h/2T_d\approx 0$, and so
for simplicity we will revert to taking $E_{0,d}= 0$ in the remainder
of the discussion.

\section{Bandlimited states}

We have provided a prescription for constructing a continuous-time
quantum hamiltonian description for any invertible classical
finite-state dynamics---turning discrete-time models into
continuous-time models.  This construction can be regarded as an
application of band\-limited interpolation theory \cite{interpol}.

\subsection{Bandlimited dynamics}\label{sec.bandlimited-dynamics}

Let us choose our unit of time such that $\tau=1$, so that our
configuration basis states are simply the states seen in the dynamics
at integer values of time starting from $\ket{0,d}$.  At a continuous
moment of time $t$ the state is
\begin{align}\label{eq.continuous-t}
  \ket{t,d} &= e^{-i\op{H}t/\hbar}\,\ket{0,d} \nonumber\\[.5em]
  &= {1\over \sqrt{N_d}}\sum_{m=0}^{N_d -1} e^{-2\pi i t
    m/N_d}\,\ket{E\! :m,d}\;,
\end{align}
which is just \eqn{inv-trans} with $t$ replacing $n$.  We can
express the continuous-time state $\ket{t,d}$ as a function of the
$N_d$ integer-time states $\ket{n,d}$ by replacing $\ket{E\! :m,d}$
with its definition \eqn{Em}:
\begin{equation}\label{eq.t-from-n}
  \ket{t,d} = \sum_{n=0}^{N_d -1} S(N_d,n-t)\:\ket{n,d}
\end{equation}
where
\begin{equation}\label{eq.S}
  S(N,u) = {1\over N} \sum_{m=0}^{N -1} e^{2\pi i m u/ N}\;.
\end{equation}
The function $S(N_d,n-t)$ equals the Kronecker delta $\delta_{n,t}$
for integer values of $t$ between 0 and $N_d-1$ but is also defined
for non-integer values.  $S(N,u)$ is a periodic version of the $\sinc$
function \cite{sinc}, which is the foundation of band\-limited
interpolation theory: $S(N,u)=1$ for integer values of $u$ that equal
0 modulo $N$ and $S(N,u)=0$ for other integer values of $u$.  In fact,
if we sum the geometric series we recover $\sinc$ times a phase for
large $N$,
\begin{equation}\label{eq.sinc-limit}
  \lim_{N\to\infty}S(N,u) = e^{i\pi u}{\sin \pi u\over \pi u}\;.
\end{equation}
A portion of the probability distribution
$\left\lvert{S(N,u)}\right\rvert^2$ is shown in
Figure~\ref{fig.prob-dist} for $N=100$ (solid).  Near its center it is
approximately gaussian (dashed).

\fign{prob-dist}{ \hfill \mbox{
    \includegraphics[width=3in]{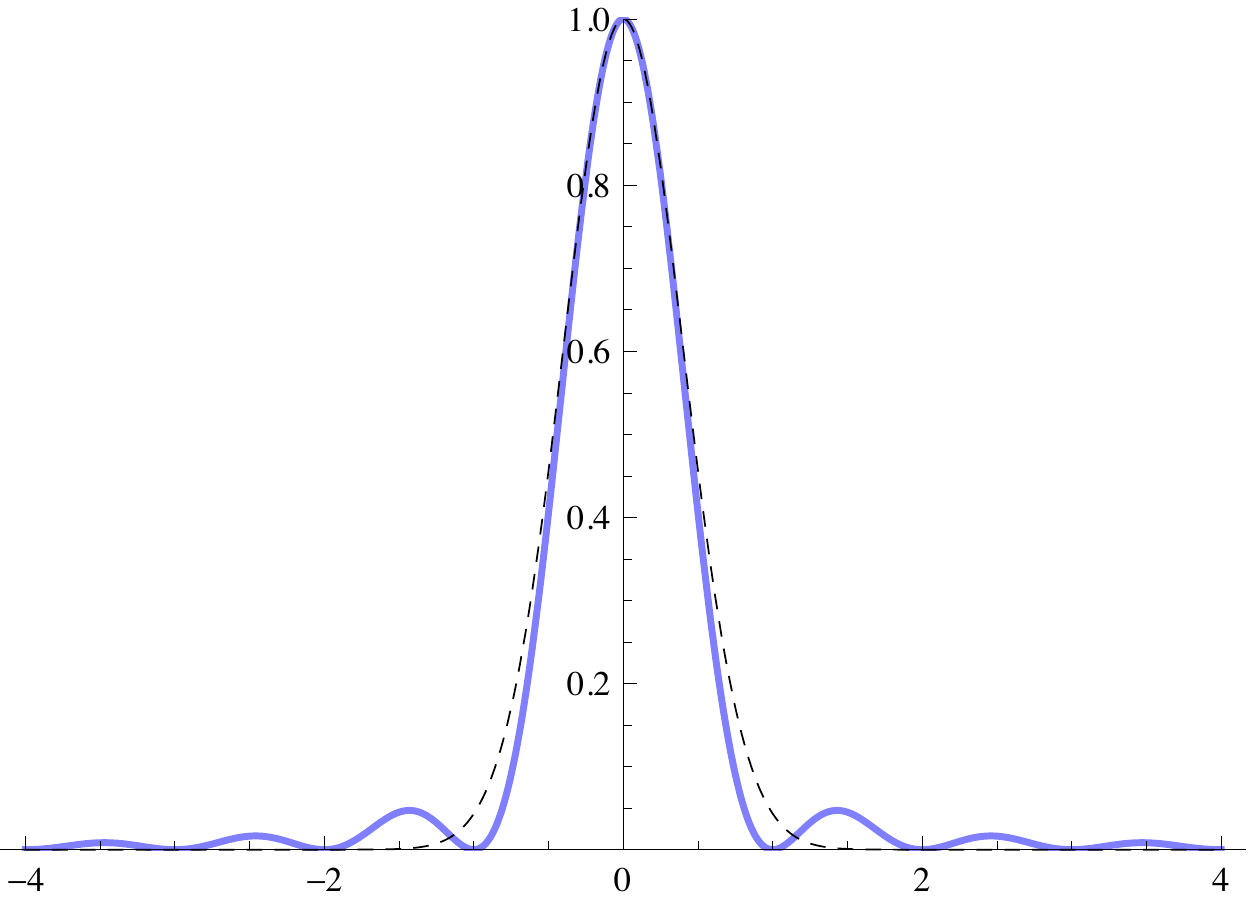}}\hfill} {
  $\left\lvert{S(N,u)}\right\rvert^2$ probability distribution (solid)
  versus normalized gaussian of unit height (dashed).}

\subsection{Reconstruction from samples}

Using $S(N,u)$, any periodic function $f(t)$ with period $T$ and a
band\-limited Fourier spectrum with $N$ frequencies can be
reconstructed from $N$ equally spaced samples.  Because of the
periodicity all frequencies must be integer multiples of $1/T$, and if
the lowest frequency is 0, then
\begin{equation}
  f(t)=\sum_{m=0}^{N-1} a_m e^{2\pi i m t/T}\;
\end{equation}
for some set of $a_m$.  Using $\tau = T/N$, $f(t)$ is also given by
\begin{equation}\label{eq.n-samples}
  f(t) = \sum_{n=0}^{N-1} f(n\tau) \: S(N,{t\over\tau}-n)\;.
\end{equation}

This is obviously true at the $N$ sample times $t=n\tau$ and so it
must be true at all times, since $S(N,{t\over\tau} - n)$ is composed
of the same frequency components as $f(t)$, and the $N$ coefficients
$a_m$ are completely determined by the values of $f(t)$ at the $N$
sample times $t=n\tau$ (in fact, the $a_m$'s are the Fourier transform
of the $f(n\tau)$'s).  If the lowest frequency is $k/T$ rather than
zero, use $S_k(N,u)=e^{2\pi i k u/N}\,S(N,u)$ instead of $S(N,u)$
above.

Thus \eqn{t-from-n} can be regarded as an exact reconstruction of a
continuous but band\-limited dynamics in Hilbert space from $N_d$
samples.  The band\-limit on the energy spectrum erases the
distinction between continuous-time and discrete-time dynamics (and
field operators \cite{tsang}), since a band\-limited periodic function
is completely determined by a finite number of sample points.

If $g(t)$ has the same period and bandwidth as $f(t)$ (perhaps with a
different lowest frequency) then \eqn{n-samples} implies
\begin{equation}\label{eq.int=sum}
{1\over T} \int_0^T \!dt\, f(t)\,g(t) = {1\over N}\sum_{n=0}^{N-1}
f(n\tau)\,g(n\tau)\;,
\end{equation}
and so a band\-limit also erases some of the distinction between
continuous and discrete analysis of the dynamics.


\section{Continuous isomorphism}

Rather than just have integer time states of a classical finite-state
dynamics correspond to integer time states of a quantum finite-energy
dynamics, we can also extend the classical finite-state dynamics to
intermediate times and have the two systems be isomorphic at all
times.

\subsection{Continuously extended dynamics}\label{sec.extended}

In classical finite-state lattice dynamics it is often useful to
imagine that, when a 1 representing a particle hops from one lattice
site to another, it moves continuously in between.  This extension of
the dynamics allows us to extend classical-mechanical conservations
associated with continuous spatial symmetries to discrete particle
motion in order to define, for example, momentum conserving lattice
gases \footnote{This idea is the basis of practical lattice gas models
  of classical hydrodynamics \cite{3dlga} and of quantum fluids
  \cite{yepez-prl} (both with continuous isotropy in the macroscopic
  limit).}.

Continuously extended lattice dynamics have a continuous evolution in
both time and space but, at every moment, only a finite amount of
state: if there are $n$ spots in space that can have a 1 or not at
integer times, there are still only $n$ spots that can at non-integer
times.  Since these $n$ bits don't change their values while they're
moving between integer locations, the non-integer-time states are
really just a fixed sequence of rearrangements of the bits of the
integer-time state.  These extra intermediate states are distinct
classically since the bits are in different places but they are
redundant informationally.

Note that a continuously extended lattice dynamics can still be
described as a repeated cycle of local updates, but in this case each
update moves a bit only infinitesimally.  After any finite interval of
time all of the bits will have moved by equivalent amounts.

\subsection{Continuously extended isomorphism}

In a continuously extended classical lattice dynamics, any unit-time
separated sequence of states provides a complete description of the
logical dynamics: since the bits of state don't change between integer
times, exactly when we sample them doesn't matter.

Similarly, any unit-time separated sequence of states from the
continuous unitary evolution \eqn{continuous-t} constitute a complete
orthonormal basis set, since \eqn{t-from-n} implies
\begin{equation}\label{eq.tt-inner}
\braket{t^\prime,d}{t,d} = S(N_d,t^\prime-t)\;.
\end{equation}
Thus we are free to define a distinguished basis at any time $t$
consisting of the unit-time separated set of $N_d$ states from the
evolution \eqn{continuous-t} that includes the current state
$\ket{t,d}$.  If we identify these basis states with corresponding
unit-time separated classical configurations, then the classical and
quantum dynamics are isomorphic at all times.


In analyzing finite-state dynamics, the $\ket{t,d}$'s act much like a
complete continuous basis since, again from \eqn{t-from-n},
\begin{equation}\label{eq.closure}
  \int_0^{T_d} \! dt \;\ket{t,d}\bra{t,d} = \sum_{n=0}^{N_d -1}
  \ket{n,d}\bra{n,d} = I \;.
\end{equation}
Moreover, the inner product \eqn{tt-inner} acts like a Dirac delta
function in an integral with a band\-limited function $f(t)$.  From
\eqn{int=sum},
\begin{equation}\label{eq.f-dirac}
  \int_0^{T_d} \! dt \; f(t) \,\braket{t^\prime,d}{t,d} =
  f(t^\prime)\;.
\end{equation}

The continuously extended isomorphism can be used to compute average
values for operators, such as momentum, defined on continuous sets of
configurations.

\section{Continuous hamiltonian}

Rather than use $N_d$ orthonormal quantum states to describe a
classical orbit with $N_d$ informationally distinct configurations, it
is sometimes convenient to use more.  In the continuous-basis limit
this yields a continuous-hamiltonian description.

\subsection{Oversampled dynamics}\label{sec.oversampled}

Suppose that, starting with a classical finite-state dynamics, we add
$M-1$ redundant intermediate-time states in the unit interval between
each pair of consecutive integer-time states.  Each orbit $d$ of the
corresponding quantum dynamics (generated by the hamiltonian
$\op{H}_M$) now visits $M N_d$ basis states rather than just the $N_d$
of the original dynamics (generated by $\op{H}_1$), and the state of
the new $\op{H}_M$ dynamics at a continuous moment of time becomes,
from \eqn{t-from-n},
\begin{equation}\label{eq.extended-state}
  \ket{t,d,M} = \sum_{k=0}^{M N_d-1} S(M N_d,k-M
  t)\:\ket{\frac{k}{M},d,M}\;,
\end{equation}
where the basis state $\ket{{k\over M},d,M}$ has been labeled by the
time $k/M$ when it is reached in an evolution starting from
$\ket{0,d,M}$.  Since this extended dynamics traverses distinct states
at a rate $\nu_M$ that is $M$ times the original rate $\nu$, it has
$M$ times the average energy.  As the number of intermediate states
added in a fixed time period goes to infinity, the hamiltonian
$\op{H}_M$ approaches a continuous hamiltonian $\op{H}_\infty$ and the
average energy of the state $\ket{t,d,M}$ goes to infinity.

\subsection{Bandlimited basis}

By putting a bandlimit on the energy spectrum of the configuration
basis states we can make the $\op{H}_M$ dynamics isomorphic to the
original $\op{H}_1$ dynamics, with the same average energy: a
bandlimit on energy can correct for an oversampling of the underlying
classical dynamics.

The Fourier transform relationship \eqn{Em} between energy eigenstates
and configurational basis states is left unchanged but we construct,
in addition, a new set of band\-limited configurations
$\ket{n,d,M}_{N_d}$ which are the Fourier transforms of the lowest
$N_d$ energy eigenstates of $\op{H}_M$,
\begin{equation}\label{eq.cutoff-n}
  \ket{n,d,M}_{N_d} = {1\over \sqrt{N_d}}\sum_{m=0}^{N_d -1}
  e^{-2\pi i n m/N_d}\,\ket{E\! :m,d,M}\;,
\end{equation}
with $n$ an integer.  These states constitute an orthonormal basis for
band\-limited superpositions of configurations.  They have the same
average energy as the configuration basis states of the $\op{H}_1$
dynamics: the amount of time $T_d$ taken for one period of the orbit
is being kept constant, and so from \eqn{eigen} the first $N_d$ energy
eigenvalues $mh/T_d$ of $\op{H}_M$ are the same as for $\op{H}_1$.

The continuous time states $\ket{t,d,M}_{N_d}$ that evolve from
$\ket{0,d,M}_{N_d}$ are given by \eqn{cutoff-n} with $n$ replaced by
$t$.  As in \eqn{t-from-n} they obey
\begin{equation}\label{eq.t-from-n-N}
  \ket{t,d,M}_{N_d} = \sum_{n=0}^{N_d -1} S(N_d,n-t)\:\ket{n,d,M}_{N_d}\;,
\end{equation}
so the evolution of band\-limited states is isomorphic with that of
$\ket{t,d}$.  Moreover, from \eqn{cutoff-n} with $n\to t$ and
expressing $\ket{E\!  :m,d,M}$ in terms of the $M N_d$ configurational
basis states using \eqn{Em},
\begin{flalign}\label{eq.cutoff-t}
  \ket{t,d,M}_{N_d} = {1\over \sqrt{M}}\sum_{k=0}^{M N_d-1}
  S(N_d,{{\textstyle {k\over M}}-t}\,)\:\ket{{k\over M},d,M}
  \\\label{eq.many-bases} = {1\over \sqrt{M}}\sum_{m=0}^{M-1}
  \sum_{n=0}^{N_d-1} \! S(N_d,n+{\textstyle {m\over
      M}}-t\,)\:\ket{n+{m\over M},d,M} &\;.
\end{flalign}
The band\-limited state is, at all times, an equally weighted
superposition of $M$ equivalent states, each of which corresponds to
the extended classical configuration at time $t$ represented in a
different unit-time separated basis.  Thus the correspondence of
$\ket{t,d,M}_{N_d}$ to classical configurations is the same as for
$\ket{t,d}$.

The state \eqn{cutoff-t} is a sum over configurations separated in
time by $du=1/M$.  If we normalize each configuration state to length
$\sqrt{M}$ instead of to length 1, this becomes delta-function
normalization in the limit $M\to\infty$ and
\begin{flalign}\label{eq.cutoff-t-lim}
  \ket{t,d,\infty}_{N_d} = \int_0^{T_d} \! du \:
  S(N_d,u-t)\:\ket{u,d,\infty}\;.
\end{flalign}
From this and \eqn{tt-inner},
\begin{equation}\label{eq.iso-equiv}
\braket{t^\prime,d,\infty}{t,d,\infty}_{N_d}=\braket{t^\prime,d}{t,d}\;,
\end{equation}
and so we can use the isomorphic $\ket{t,d}$ states to determine
amplitudes in the continuous configuration basis.







\section{Particle motion}\label{sec.particle-motion}


A classical finite-state lattice dynamics is naturally described as a
repeating sequence of invertible gate operations \cite{crystalline}.
In mapping this onto a quantum dynamics, the classical model can be
implemented isomorphically as a sequence of local unitary operations.

Fundamental physics is, however, normally described as particle
dynamics.  To make contact with this viewpoint we can recast
finite-state lattice dynamics as particle mechanics, {\em following
  the motions of individual 1's as if they were distinguishable
  particles}.

\subsection{Single particle}\label{sec.discrete-motion}

Consider a classical lattice dynamics in which a single particle,
represented by a 1, hops in the $+x$ direction from lattice site to
adjacent lattice site at a constant rate, with average speed $v=1$.
The motion is periodic in space, traversing $N$ lattice sites in a
distance $L$ before repeating.  At $t=0$ the particle is at $x=0$.

For this classical evolution, we can take the state of the system to
be the integer position $n$ of the 1 at integer time $n$.  In an
isomorphic $\op{H}_1$ quantum evolution, the distinct classical
configurations become integer-position basis states $\ket{n}$.  From
\eqn{t-from-n} we get a description of intermediate configurations in
terms of integer-time ones,
\begin{equation}\label{eq.interpol-x}
  \ket{x} = \sum_{n=0}^{N -1} S(N,n-x)\:\ket{n}\;,
\end{equation}
where $\ket{x}$ is the configuration obtained by evolving for a time
$t=x/v$ from the configuration $\ket{0}$.  We identify the non-integer
$\ket{x}$ with the non-integer positions of the continuously extended
dynamics.


In the quantum description of a classical particle at a non-integer
position $vt$ modulo $L$, there is some amplitude for the particle at
more than one integer position.  From \eqn{tt-inner} and using
\eqn{iso-equiv} we can interpret
\begin{equation}\label{eq.psi}
\psi(x,t)=\braket{x}{vt}=S(N,x-vt)
\end{equation}
to be the amplitude to find the particle at any continuous position
$x$ at time $t$, and compute the average momentum directly from
$\psi(x,t)$.

Alternatively, we can instead start with an infinite-dimensional
quantum hamiltonian that generates a continuous shift in space in the
$+x$ direction at speed $v$:
\begin{equation}\label{eq.h-infty}
  \begin{array}{ccc}
    \op{H}_\infty=v\tts\op{p}, & \mbox{with} & \op{p}=-i\hbar{\partial
      \over \partial x}\;.
  \end{array}
\end{equation}
The direction of the shift is apparent from noting that
$\op{H}_\infty\psi=i\hbar\,\partial\psi/\partial t$ implies
$\psi(x,t)=\psi(x - vt,0)$ \footnote{For motion in the $-x$ direction
  we would use $\op{H}_\infty=-v\op{p}$ instead.  To represent the
  direction explicitly in the state we would multiply it by $\ket{+}$
  or $\ket{-}$ and let $\op{H}_\infty=\sigma_z v\op{p}$.}.  Now we can
make this dynamics isomorphic to the $\op{H}_1$ discrete shift by
band\-limiting the initial state so that the evolution only traverses
$N$ distinct states in the width $L$.  Then from \eqn{cutoff-t-lim}
the state corresponding to a classical particle at position $vt$ in
the position basis is $\psi(x,t)=S(N,x-vt)$, with energy $E=hN/2T$
just as in $\op{H}_1$.  From \eqn{h-infty},
\begin{equation}\label{eq.p}
p = {E\over v} = {h\over 2\lambda}\;,
\end{equation}
where $\lambda=L/N$.  The state $S(N,x-vt)$ achieves a general bound
$\lambda \ge h/2p$ on the average separation of distinct states of a
moving particle \cite{first-mom}.

Since this description applies to any particle shifting uniformly in a
lattice dynamics, \eqn{p} gives the corresponding momentum.  Of course
only lattice update operations that actually move a particle
contribute to the shift-energy $E=vp$ portion of its total
energy \footnote{To make updates local, we can use a {\em partitioning
    dynamics,} in which each particle motion involves a single update
  that changes both old and new positions at once
  \cite{crystalline,bcq}.}.

\subsection{Classical mechanics}

We can often consider a classical lattice-gas dynamics to be a
discrete-time sampling of an idealized classical-mechanical particle
dynamics \cite{fredkin-bbm,soft-spheres} that obeys Hamilton's
equations,
\begin{equation}\label{eq.hamiltons}
{\partial H \over \partial q_j} = -{dp_j \over dt}\;, \qquad {\partial
  H \over \partial p_j} = {dq_j \over dt}\;.
\end{equation}
To make the lattice dynamics run faster by a factor $\kappa$ we reduce
the interval between the discrete events, $\tau\to\tau/\kappa$.  From
\eqn{hamiltons}, this can be accomplished by letting $H\to \kappa H$,
which is exactly the energy scaling required by \eqn{E}.

We can't just rescale $\tau$ arbitrarily while keeping the $p_j$'s and
$q_j$'s unchanged, however, because particle velocities are limited by
the speed of light.  We can, instead, run the dynamics faster by
putting the discrete events closer together in both time and space,
leaving velocities unchanged.  If the distance between events
$\lambda\to\lambda/\kappa$, then the scale of the $p_j$'s must be
multiplied by $\kappa$ to get an overall scaling of $H$ by $\kappa$ in
\eqn{hamiltons}.  This is exactly the momentum scaling required by
\eqn{p}.

\subsection{Indistinguishable particles}




Treating 1's in a classical finite-state lattice dynamics as
distinguishable particles---and keeping track of the discrete position
and velocity of each 1---dramatically over-represents the number of
distinct states: all states with the same spatial pattern of 1's and
velocities correspond to a single state of the original lattice model.
We can fix this over-representation in a quantum description of the
distinghishable particle dynamics by merging equivalent states, adding
them together to form new {\em occupation number} basis states, and
using only these to describe the evolution.  If we antisymmetrize each
sum under particle interchange, the new basis states will each have at
most one 1 with a given position and velocity---we can symmetrize
instead to allow more \cite{ilga}.

To describe a dynamics in which the number of ones changes with time,
we can use creation and annihilation operators to add and remove
particles from the state, while maintaining symmetrization.  These
field operators inherit fermionic or bosonic commutation rules from
the symmetrization \footnote{For example, if particle labels are
  generated sequentially as particles are created, then interchanging
  the order in which two particles are added to an antisymmetrized
  state is equivalent to interchanging their particle labels, and so
  creation operators must anticommute \cite{ziman}.}.  As we see from
\eqn{interpol-x} (or from \eqn{t-from-n-N} for $\op{H}_\infty$), a
finite set of band\-limited basis states allows a particle to be added
centered at any continuous position in space.  In one dimension with
one velocity, for example, the creation operator
$\op{\Psi}^\dagger(x)$ for any $x$ is a superposition of the creation
operators $\op{\Psi}^\dagger(n)$ for integer positions $n$,
\begin{equation}
\op{\Psi}^\dagger(x) = \sum_{n=0}^{N-1} S(N,n-x)\,
\op{\Psi}^\dagger(n)\;.
\end{equation}

Of course nothing essential is gained by using a continuous space and
time description, since a band\-limited continuous state is completely
determined by its values at discrete positions and times.  Similarly,
nothing essential is gained by introducing fermionic field operators:
there would be no need to maintain the antisymmetry of equivalent
states if the original dynamics were described isomorphically in terms
of local unitary operations \footnote{This should apply equally to
  quantum lattice gas simulations of non-classical systems.}.





%



\section{Uncertainty}

The particle described by \eqn{psi} moves at a constant speed and is
localized to a single position basis state of a finite-dimensional
basis at all times ({\em cf.}~\cite{braginsky}).  This in no way
conflicts with the uncertainty relations of quantum mechanics, which
can be regarded as bounds on representing information using limited
band\-width.

\subsection{Bandwidth bounds}

Constraints on time or position determine the minimum width of the
energy or momentum eigenfrequency distribution needed to describe a
state that meets the constraints.  In the usual uncertainty bounds we
also associate a width with the time or position amplitude
distribution \cite{uffink}, but in general other constraints on time
or position can be used to determine a minimum width of energy or
momentum eigenfrequencies.

For example, suppose we have an exactly periodic evolution with period
$T$.  The state at time $t$ can be written
\begin{equation}\label{eq.kett}
  \ket{t} = \sum a_n e^{- 2\pi i \nu_n t}\,\ket{E_n}\;.
\end{equation}
Exact periodicity requires that each $\nu_n=E_n/h$ be an integer
multiple of $1/T$.  If this evolution passes through $N$ mutually
orthogonal states, then the superposition must involve at least $N$
different $\ket{E_n}$'s (since you can't construct $N$ distinct states
out of fewer than $N$ distinct states).  Moreover, there must also be
at least $N$ distinct frequencies (since groups of $\ket{E_n}$'s with
the same frequency act like a single eigenstate in the construction).
To have $N$ distinct frequencies that are integer multiples of $1/T$,
the band\-width $B$ (highest frequency in the superposition minus
lowest) must obey
\begin{equation}\label{eq.B}
  B \ge {N-1\over T}\;.
\end{equation}
This is a version of the bandwidth-time theorem of communication
theory \cite{slepian-bandwidth}.  If we let $\tau=T/N$ be the average
time between distinct states, we see that this is also a version of
the time-energy uncertainty relation, using $B$ directly rather than
some other measure of the width of the energy eigenfrequency
distribution.  The definition \eqn{inv-trans} achieves this bound.

\subsection{Second-moment bounds}

In constructing uncertainty bounds, the standard deviation of the
eigenfrequency distribution is traditionally chosen to measure its
width.  This choice reflects both familiarity from statistics and (for
position and momentum) a simple connection between the commutation
relation and the standard-deviation bound \cite{robertson}.  This
choice is often divergent, however, and so fails to provide a useful
bound \cite{uffink}.  This is true in our case.

Consider the band\-limited state $\braket{x}{\xb} = S(N,x-\xb)$
centered at $\xb$.  Limiting ourselves to spatial frequencies $m/L$
with $m$ ranging from 0 to $N-1$, this state has the least possible
information about what the momentum is, since all momentum eigenstates
in the allowed range have equal amplitude.  Correspondingly we might
expect the position to be as well-defined as possible, given the
limited band\-width.  It is clear from Figure~\ref{fig.prob-dist} that
the position localization of the probability distribution
$\av{S(N,x-\xb)}^2$ is similar to that of a gaussian (dotted line).
This is not apparent in the mean square position deviation, however,
which can be estimated for large $N$ using \eqn{sinc-limit} as
\begin{equation}
  \bk{(x-\bar{x})^2}\approx\int_0^N \!\!(x-\bar{x})^2{\sin^2 \pi
    (x-\bar{x})\over \pi^2(x-\bar{x})^2}\,dx={N\over 2\pi^2}\;,
\end{equation}
which diverges as $N\to\infty$ ({\em i.e.,} on an infinitely wide
space) \footnote{If $\xb$ is near the middle of the periodic space,
  then the wavefunction goes to 0 at the boundary as $N\to\infty$ and
  so the usual uncertainty relations apply \cite[Eqn.\ 22]{pb}.}.
Thus $S(N,x-\xb)$, which is perfectly distinct from a unit shift of
itself, is not localized at all on the infinite line if we use the
traditional second-moment measure of the width of the distribution.
The unit-height gaussian, which looks so similar in the figure, has a
mean square deviation of $1/2\pi$.  Other measures of the width have
been proposed that avoid this disparity \cite{uffink}.


\subsection{First-moment bounds}

For our purposes, a much better measure of the width of the
eigenfrequency distribution is twice the average half-width:
$2(\bar{\nu} - \nu_0)$.  Here $\bar{\nu}$ is the average frequency
({\em e.g.,} $E/h$) and $\nu_0$ the lowest frequency used ({\em e.g.,}
$E_0/h$).  In general \cite{first-mom},
\begin{equation}\label{eq.av-min}
  2(\bar{\nu} - \nu_0) \ge B_\smin\;,
\end{equation}
where $B_\smin$ is the minimum band\-width compatible with the
temporal or spatial constraints on the system.

For example, if $\taumin$ is the minimum separation in time between
two mutually orthogonal states in the evolution, then the minimum
band\-width needed is $B_\smin=1/2\taumin$: there must be at least two
distinct frequencies and they must be separated by at least half of
$1/\taumin$.

The $B_\smin=1/2\taumin$ bound \eqn{av-min} is only achieved by the
energy \eqn{EN} for $N=2$.  For $N\gg2$, the energy \eqn{EN} is about
twice as great as allowed by this bound.  There is, however, the
additional band\-width constraint \eqn{B} required to have $N$
distinct states in period $T$.  The energy \eqn{EN} achieves
\eqn{av-min} with this constraint.

\subsection{Uncertain states}

We have seen examples where a quantum hamiltonian describes a
classical finite-state dynamics, but also makes extra distinctions not
present in the original dynamics: A many particle hamiltonian that
keeps track of which identical 1-bit is where.  A continuous-shift
hamiltonian that adds distinct states between the discrete time steps.

We can eliminate over-representation and make the dynamics isomorphic
to the original by adding together equivalent configurations with
equal weight to construct truly distinct basis states.  Starting from
these, equivalent configurations will always have equal probability:
{\em equivalence is represented as uncertainty} \footnote{Ignorance of
  differences between equivalent states doesn't count toward entropy,
  which is one reason quantum probabilities must be kept separate from
  ordinary ones \cite{bcq}.}.

In the construction of the occupation number basis states for
identical 1-bits, a symmetrized or antisymmetrized state represents
equivalent states as being equally probable.  In the case of
over-representation of intermediate states, constructing a basis
without the high frequency information needed to represent
intermediate details also merges equivalent states \eqn{cutoff-t},
making them equally probable.

The continuous-hamiltonian representation of a discrete shift is an
interesting limiting case of representing equivalence as uncertainty.
A band\-limit with $N$ distinct states yields \eqn{many-bases} for
finite $M$.  For a state centered at $t=x/v$ and $M\to\infty$ this
becomes
\begin{align}\label{eq.unit-sup}
  \ket{x,\infty}_N &= \int_0^1 du\;\left(\sum_{n=0}^{N-1}
  S(N,n+u-x)\,\ket{n+u,\infty}\right)\;,
\end{align}
which is a uniform superposition of all the equivalent ways to
represent a classical particle at position $x$ if only $N$
equally-spaced positions are distinct.

The tradeoff between band\-width and minimum separation in space
determines the minimum uncertainty volume of phase space needed to
represent each distinct state \footnote{In a periodic space of length
  $L$, momentum eigenfrequencies must be integer multiples of $1/L$.
  Thus to represent $N$ distinct states a band\-width $B\ge (N-1)/L$
  is needed, and so the frequency-space volume per distinct state is
  $BL/N \ge (N-1)/N$ (uncertainty tradeoff for $B$ vs. $L/N$).}, and
this is achieved by $\ket{x,\infty}_N$.






\section{Discussion}\label{sec.discussion}








Classical finite-state dynamics that are invertible can be mapped
isomorphically onto the discrete time behavior of finite-energy
quantum dynamics.  A quantum evolution mapping an infinite number of
distinct states into a finite time period would have an infinite
average energy.

Quantum-classical isomorphism challenges conventional wisdom about
essential differences between quantum and classical systems: identical
particles, amplitudes, frequencies, complementarity and uncertainty
all play essential roles in describing and analyzing classical
finite-state dynamics using continuous language.

Quantum-classical models also shed light on the foundations of
classical mechanics.  They provide a quantum substrate where
interesting classical behavior arises without approximation or
decoherence.  Physically meaningful energy and momentum scales are
defined directly by the separation of classical events in time and
space.


Finally, quantum-classical isomorphism may be helpful in understanding
and teaching quantum mechanics.  Just as it is useful to study
classical information and classical computation as a preliminary to
studying their quantum counterparts, it seems useful to study other
aspects of the machinery and concepts of quantum mechanics in a
simplified classical setting.


\end{document}